\documentclass[prl,amsmath,groupedaddress,floatfix,twocolumn]{revtex4-1}

\usepackage{graphicx}
\usepackage{amssymb}
\usepackage{longtable}
\usepackage{rotating}
\usepackage{array}
\usepackage{multirow}
\usepackage{makecell}
\usepackage{booktabs}
\usepackage{color}

\begin{document}

\title{Perturbative approach to electrocaloric effects}

\author{M\'onica Graf$^{1}$ and Jorge \'I\~niguez$^{1,2}$}

\affiliation{
  \mbox{$^{1}$Materials Research and Technology Department,
    Luxembourg Institute of Science and Technology (LIST),} \mbox{Avenue
    des Hauts-Fourneaux 5, L-4362 Esch/Alzette,
    Luxembourg}\\
 \mbox{$^{2}$Department of Physics and Materials Science, University
   of Luxembourg, Rue du Brill 41, L-4422 Belvaux, Luxembourg} }
  
\begin{abstract}
We show that the electrocaloric (EC) effect -- e.g., the temperature
change experienced by an insulator upon application of an electric
bias -- lends itself to a straightforward interpretation when
expressed as a Taylor series in the external field. Our formalism
explains in a unified and simple way the most notable small-field
effects reported the literature -- e.g., the so-called {\em normal}
(increase of temperature under bias, as usually found in
ferroelectrics) and {\em inverse} (decrease of temperature, as e.g. in
antiferroelectrics) EC responses -- and clarifies their physical
interpretation. We also discuss in detail second-principles
simulations for prototype ferroelectric PbTiO$_{3}$, explicitly
evaluating subtle predictions of the theory, such as the occurrence of
competing contributions to the EC response.
\end{abstract}

\maketitle

Electrocaloric (EC) effects hold the promise of an ecofriendly
alternative to refrigeration, one of the most energy-consuming
activities today and in the foreseeable future. EC effects are strong
in ferroelectric (FE) and antiferroelectric (AFE) materials because of
their anomalously large dielectric responses and the field-driven
phase transitions that can be easily induced near the Curie point
\cite{kutnjak15}.

The thermodynamic description of the EC cooling cycle is well
established \cite{kutnjak15}. The key step is the
  adiabatic temperature change, which can be expressed as
\begin{equation}
  \begin{split}
    \Delta T(\alpha) & =- \int_{0}^{E_{\alpha}} \frac{T}{C_{\bf E}}
  \left(\frac{\partial S}{\partial E'_{\alpha}}\right)_{T}
  dE'_{\alpha} \\
 & = - \int_{0}^{E_{\alpha}} \frac{T}{C_{\bf E}}
  \left(\frac{\partial P_{\alpha}}{\partial T}\right)_{\bf E}
  dE'_{\alpha} \\
 & = - \int_{0}^{E_{\alpha}} \frac{T}{C_{\bf E}} \pi_{\alpha}
  dE'_{\alpha}  \; ,
  \end{split}
  \label{eq:deltaT}
\end{equation}
where the intregral runs from zero to a final electric field
$E_{\alpha}$, applied along the $\alpha$ Cartesian direction. $S$, $T$
and $C_{\bf E}$ are, respectively, the entropy, temperature and
constant-field heat capacity. $P_{\alpha}$ is the polarization
component conjugate to the applied field $E_{\alpha}$, and its
$T$-derivative is the pyroelectric coefficient $\pi_{\alpha}$. Note
that we have used the Maxwell relation between the field derivative of
the entropy and the pyroelectric vector, which is expected to hold for
ergodic materials. (Thus, the formalism introduced here is, in
principle, not applicable to relaxor ferroelectrics.) Note also that
the repeated $\alpha$ index does {\em not} involve an implicit sum; we
write this $\alpha$, on both sides of the equation, to emphasize the
fact that the EC response can be anisotropic.

Also important is the isothermal entropy change
\begin{equation}
  \Delta S(\alpha) = \int_{0}^{E_{\alpha}} \pi_{\alpha} dE'_{\alpha}  \; ,
  \label{eq:deltaS}
  \end{equation}
which quantifies the amount of heat the EC material will exchange with
the object to be cooled. The product $\Delta T(\alpha)\Delta
S(\alpha)$ is usually taken as the figure of merit for EC cooling
performance.

These expressions are deceivingly simple, as all the quantities in the
integrands depend on both field and temperature (itself field
dependent, by virtue of the EC effect). Further, they may present
complex behaviors (in particular, discontinuities) across field- and
temperature-driven phase transitions.

There are well-known strategies to solve these integrals
self-consistently \cite{kutnjak15}; this will not be our focus
here. Rather, we want to examine the ingredient that has the greatest
influence in the basic features (magnitude and sign) of the EC
temperature and entropy changes, namely, the pyroelectric
vector ${\boldsymbol \pi}$. Indeed, ${\boldsymbol
  \pi}$ fully determines $\Delta S({\alpha})$; further, of the
quantities contributing to $\Delta T({\alpha})$, ${\boldsymbol \pi}$
is likely to be the most sensitive to an electric bias and, more
importantly, it is the only one whose sign is not defined. (Both $T$
and $C_{E}$ are always positive.)

Interestingly, we can rewrite Eq.~(\ref{eq:deltaT}) in differential
form as
\begin{equation}
  \frac{dT(\alpha)}{dE_{\alpha}} = - \frac{T}{C_{\bf E}} \left(
  \frac{\partial S}{\partial E_{\alpha}} \right)_{T} \; .
\end{equation}
This expression is the basis for the physical interpretation of EC
effects \cite{kutnjak15}. For example, it is experimentally observed
that, in ferroelectrics, the application of an electric field usually
results in a positive temperature change, which is attributed to the
fact that electric fields create order (reduce the entropy) in such
systems \cite{kutnjak15}. In contrast, negative values of $\Delta T$
have been observed in antiferroelectrics
\cite{pirc14,geng15,novak18,li20}, or when a field is applied
perpendicular to the polarization of a ferroelectric phase
\cite{grunebohm18}, which is compatible with the intuitive notion
that, in such cases, the electric field will destabilize the
equilibrium state and cause disorder (increase the entropy). These
diverse observations have been reproduced by atomistic simulations
\cite{prosandeev08,marathe16,lisenkov16,marathe17,glazkova17,jiang17,kingsland18}
and explained by the corresponding Landau phenomenological theories
\cite{kutnjak15,jiang17,pirc14,jiang18}, suggesting that we understand
EC effects quite well.

However, we think the situation is not fully satisfactory, for one
main reason: The theoretical treatments in the literature are case
specific, and we lack a unified and simple picture revealing the
similarities and differences among the various known effects. Also, we
find that, while intuitively appealing, some frequent assumptions
(i.e., that electric fields cause disorder in antiferroelectrics) are
somewhat vague, and we miss a formalism that allows us to interpret
the observed behaviors (and eventually think about new ones) in a more
rigorous manner. Here we address these issues.

{\sl Formal considerations}.-- Let us focus on how $\pi_{\alpha}$
controls $\Delta T(\alpha)$.  For the sake of simplicity, we work with
an approximate version of the adiabatic temperature change
\begin{equation}
  \Delta T(\alpha) \approx -\frac{T^{(0)}}{C_{\bf E}^{(0)}}
  \int_{0}^{E_{\alpha}} \pi_{\alpha}(T^{(0)}) dE'_{\alpha} \; ,
  \label{eq:approxdeltaT}
\end{equation}
where the ``0'' superscript marks values at zero applied field. This
approximate expression captures the EC effect as obtained when we
neglect the field and temperature dependences of $T$ and $C_{\bf E}$,
as well as the variation of $\pi_{\alpha}$ upon EC heating or cooling;
these are common approximations in the EC literature. (Unless we work
at low temperatures, we can safely assume $|\Delta T(\alpha)|\ll
T^{(0)}$, as the largest measured EC effects are typically below 20~K
\cite{moya14}. Also, we can expect relatively small variations in
$C_{\bf E}$ except in the close vicinity of phase transitions; more on
this below.)

Let us examine Eq.~(\ref{eq:approxdeltaT}) analytically. We start by
writing ${\bf P}$ as a Taylor series in ${\bf E}$,
\begin{equation}
  P_{\mu} = P^{(0)}_{\mu} + \epsilon_{0} \sum_{\beta}
  \chi^{(0)}_{\mu \beta}E_{\beta} +\epsilon_{0} \sum_{\beta\gamma}
  \chi^{(1)}_{\mu\beta\gamma} E_{\beta}E_{\gamma} + ... \; ,
\end{equation}
where ${\bf P}^{(0)}$ is the spontaneous polarization, ${\boldsymbol
  \chi}^{(n)}$ is the $n$th-order dielectric susceptibility tensor,
and $\epsilon_{0}$ is the vacuum permittivity. For convenience, we
work with a Cartesian coordinate system with one axis parallel to
$\alpha$, the direction of the applied field. Hence, we have
$E_{\beta}= \delta_{\beta\alpha}E_{\alpha}$, where
$\delta_{\beta\alpha}$ is the Kronecker delta; we thus obtain the
$\alpha$ component of the polarization as
\begin{equation}
  P_{\alpha} = P^{(0)}_{\alpha} + \epsilon_{0} \chi^{(0)}_{\alpha
    \alpha}E_{\alpha} +\epsilon_{0} \chi^{(1)}_{\alpha\alpha\alpha}
  E_{\alpha}^{2} + ... \; ,
\label{eq:p}
\end{equation}
where only the $\alpha$-diagonal tensor elements appear.  This
expression for $P_{\alpha}$ is general and can be used to describe any
phase of an insulating material, be it FE, AFE, paraelectric (PE), or
simply dielectric.

By taking the $T$-derivative of Eq.~(\ref{eq:p}) at constant field, we
obtain
\begin{equation}
  \pi_{\alpha} = \pi^{(0)}_{\alpha} + \pi^{(1)}_{\alpha
    \alpha}E_{\alpha} + \pi^{(2)}_{\alpha\alpha\alpha}E_{\alpha}^{2} +
  ... \; ,
  \label{eq:pi}
\end{equation}
where the ${\boldsymbol \pi}^{(0)}$ vector captures
the $T$-dependence of the spontaneous polarization ${\bf
  P}^{(0)}$. For $n\geq 1$, the ${\boldsymbol \pi}^{(n)}$ tensors
account for the field-induced pyroelectric effect, with
\begin{equation}
{\boldsymbol \pi}^{(n)} = \epsilon_{0} \left(\frac{\partial
  {\boldsymbol \chi}^{(n-1)}}{\partial T}\right)_{\bf E} \; ,
\label{eq:pyro}
\end{equation}
where the pyroelectric tensor of $n$th-order in the
  field series depends on the susceptibility of ($n-1$)th-order.
Using Eq.~(\ref{eq:pi}), we resolve Eq.~(\ref{eq:approxdeltaT}) and
obtain
\begin{equation}
  \begin{split}
    \Delta T(\alpha) \approx & -\frac{T^{(0)}}{C_{E}^{(0)}} \left(
    \pi^{(0)}_{\alpha} E_{\alpha} +
    \frac{1}{2}\pi^{(1)}_{\alpha\alpha}E_{\alpha}^{2} \right. \\
    & \left.  +
    \frac{1}{3}\pi^{(2)}_{\alpha\alpha\alpha}E_{\alpha}^{3} + ... \right)\\
    = & \Delta T^{(1)}(\alpha) + \Delta T^{(2)}(\alpha) + \Delta
    T^{(3)}(\alpha) + ...  \; , \\
  \end{split}
 \label{eq:newdeltaT} 
\end{equation}
where we define $\Delta T^{(n)}(\alpha)$ as the $n$th-order
contribution to the adiabatic temperature change. Also, here it is
implicitly assumed that the pyroelectric coefficients are evaluated at
$T^{(0)}$. Let us now see how this expression allows us to understand
all the known EC effects in FE and AFE compounds for
  small applied electric fields. Our conclusions are summarized in
Table~\ref{tab:summary}.

\begin{table}[t!]
  \caption{Summary of EC effects in FE and AFE materials. Shown are
    the expected signs of the linear and quadratic contributions to
    $\Delta T({\alpha})$ ($\Delta T^{(1)}({\alpha}$) and $\Delta
    T^{(2)}({\alpha})$ columns, see text), as well as the sign of the
    experimentally measured $\Delta T$ (``exp.'' column).}
\vskip 1mm  
  \begin{tabular*}{0.9\columnwidth}{@{\extracolsep{\fill}}ccccc}
\hline\hline\\[-1.0\medskipamount]
    Case, ${\bf P}^{(0)}$ & ${\bf E}$ & $\Delta T^{(1)}(\alpha)$ &
    $\Delta T^{(2)}(\alpha)$ & exp.  \\[1ex]
\hline\\[-1.0\medskipamount]
\makecell{PE ($T>T_{\rm C}$)\\${\bf P}^{(0)}={\bf 0}$} & any & $0$ & $>0$ & $>0$ \\[2ex]
\makecell{AFE ($T<T_{\rm C}$)\\${\bf P}^{(0)}={\bf 0}$} & any & $0$ & $<0$ & $<0$ \\[2ex]
\multirow{3}{*}{\makecell{FE ($T<T_{\rm C}$)\\$(0,0,P^{(0)}_{z}>0)$}}
& $E_{x}$ & $0$ & $<0$ & $<0$ \\[0.5ex]
& $E_{z}>0$ & $>0$ & $<0$ & $>0$\\[0.5ex]
& $E_{z}<0$ & $<0$ & $<0$ & $<0$\\[0.5ex]
\hline\hline
  \end{tabular*}
  \label{tab:summary}
\end{table}

For the sake of concreteness, here we discuss FE and AFE materials
with an isotropic (e.g., cubic) high-temperature PE phase, as is the
case of perovskite oxides (e.g., PbTiO$_{3}$ or PbZrO$_{3}$), noting
that our arguments can be generalized.

Let us begin by discussing the EC response above the Curie temperature
($T_{\rm C}$) for a material that can be either FE or AFE. In this
case we do not have any spontaneous polarization (${\bf
  P}^{(0)}={\boldsymbol \pi}^{(0)}={\bf 0}$); hence, $\Delta T(\alpha)
\approx \Delta T^{(2)}(\alpha)$, which is dominated by the
lowest-order field-induced pyroelectric effect,
${\boldsymbol\pi}^{(1)}$. As we know from Eq.~(\ref{eq:pyro}),
${\boldsymbol\pi}^{(1)}$ is just the $T$-derivative of the linear
dielectric susceptibility ${\boldsymbol\chi}^{(0)}$.

\begin{figure*}[t!]
  \includegraphics[width=\textwidth]{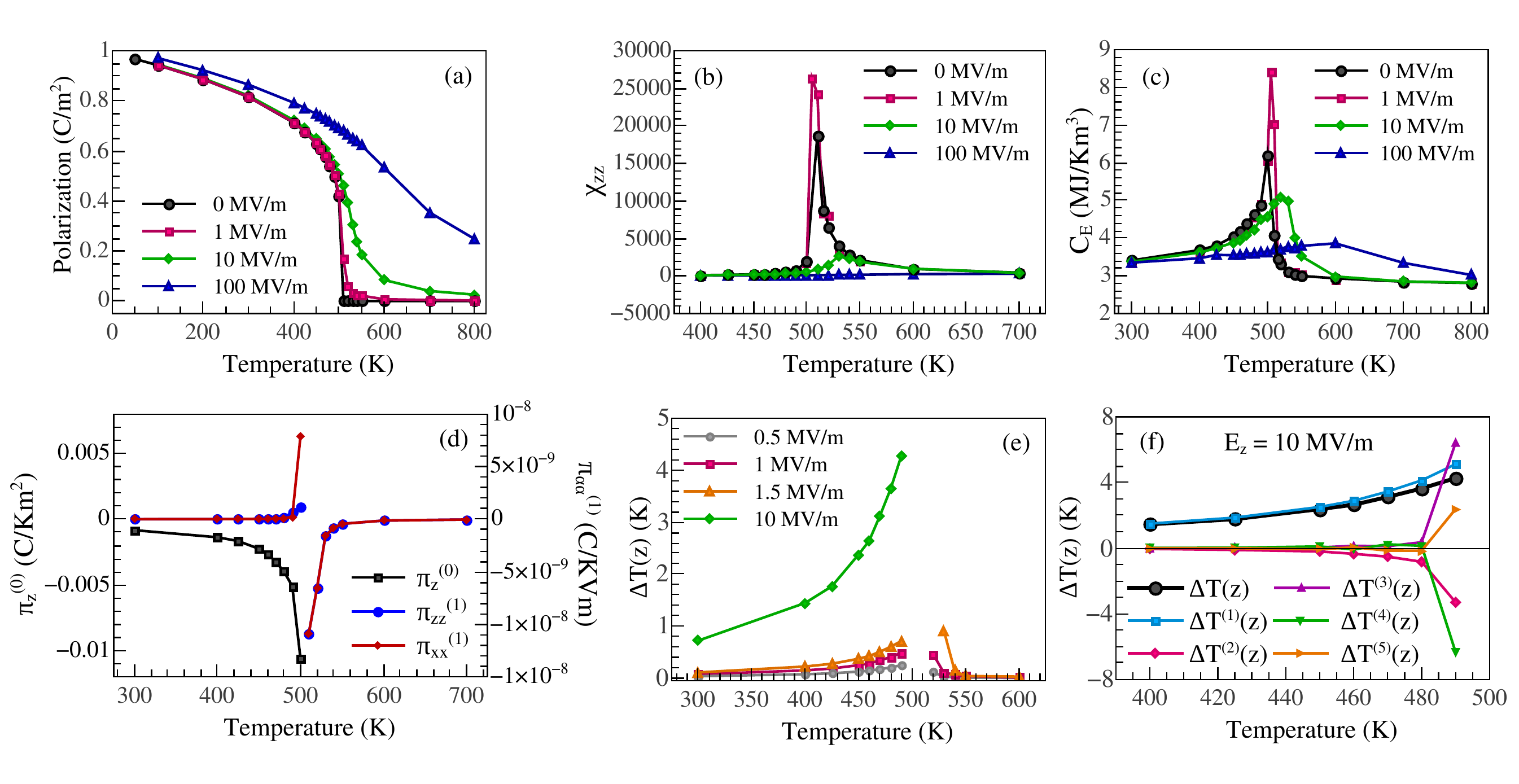}
 \vskip -4mm 
  \caption{Summary of the second-principles results obtained for
    PTO. We show the temperature and field dependence of ${\bf P}$
    (panel~(a)), ${\boldsymbol \chi}$ (panel~(b)), and specific heat
    $C_{\bf E}$ (panel~(c)). Panel~(d) shows the $T$-dependence of the
    $\pi^{(0)}_{z}$, $\pi^{(1)}_{xx}$, and $\pi^{(1)}_{zz}$ components
    of the pyroelectric tensors; they correspond to the case in which
    we have a positive spontaneous polarization along $z$, as
    discussed in the text. Also shown is $\Delta T(z)$ obtained for
    different electric fields (panel~(e)), as well as the result for
    $E_{z} = 10$~MV/m decomposed in the different $\Delta T^{(n)}(z)$
    contributions (panel~(f)). In panels (e) and (f): no results are
    shown for $T\gtrsim T_{\rm C}$ whenever the applied ${\bf E}$
    induces a transition.}
  \label{fig:pto}
\end{figure*}

As it is well-known, both FE and AFE phase transitions are
characterized by a dielectric anomaly at zero field, i.e., a maximum
of ${\boldsymbol \chi^{(0)}}$ at $T_{\rm C}$. To fix ideas, we can
imagine that all the diagonal components of ${\boldsymbol \chi^{(0)}}$
follow a Curie-Weiss law approximately. (Our simulation results for
PbTiO$_{3}$ -- see Fig.~\ref{fig:pto} -- are a representative case.)
This maximum controls the $T$-dependence of ${\boldsymbol \chi^{(0)}}$
in a wide range around $T_{\rm C}$, yielding
$\pi^{(1)}_{\alpha\alpha}>0$ for $T<T_{\rm C}$ and
$\pi^{(1)}_{\alpha\alpha}<0$ for $T>T_{\rm C}$. Hence, in particular,
the PE phase of all FEs and AFEs is characterized by
$\pi^{(1)}_{\alpha\alpha}<0$; according to Eq.~(\ref{eq:newdeltaT}),
this should result in $\Delta T(\alpha)>0$ for all $\alpha$
directions, as it is indeed observed.

Interestingly, the situation is rather similar for an AFE state: we
still have ${\bf P}^{(0)}={\boldsymbol \pi}^{(0)}={\bf 0}$ and $\Delta
T(\alpha) \approx \Delta T^{(2)}(\alpha) \propto
\pi^{(1)}_{\alpha\alpha}$. However, now we have $T<T_{\rm C}$ and, as
mentioned above, $\pi^{(1)}_{\alpha\alpha}>0$. Hence, we expect
$\Delta T(\alpha)<0$, in agreement with the experimental
observations.

Note that in all the above cases the EC temperature change does not
depend on the sign of the applied field, a feature that is expected
from the symmetry of PE and AFE states, and which we readily obtain
from our formalism.

Suppose now that we are in a FE phase. Without loss of generality, we
assume that ${\bf P}^{(0)}$ is parallel to the $z$ Cartesian
direction, with $P^{(0)}_{z}>0$ and $P^{(0)}_{x}=P^{(0)}_{y}=0$. In
this case, the ${\boldsymbol \pi}^{(0)}$ vector has
one non-zero component ($\pi^{(0)}_{z}<0$) and two null ones
($\pi^{(0)}_{x}=\pi^{(0)}_{y}=0$). Further, we have
$\pi^{(1)}_{\alpha\alpha}>0$ for all $\alpha$ (from the Curie-Weiss
law and the fact that $T<T_{\rm C}$).

Let us imagine we apply a field $E_{x}$ along the $x$ Cartesian
direction, thus perpendicular to ${\bf P}^{(0)}$. This case is exactly
analogous to the AFE state discussed above: according to
Eq.~(\ref{eq:newdeltaT}), the response is controlled by
$\pi^{(1)}_{xx} > 0$. Hence, exactly as in the AFE state, we expect an
inverse EC response with $\Delta T(x)<0$, an experimentally-observed
behavior that might seem surprising \cite{grunebohm18,perantie10}, but
is readily obtained and explained within our formalism.

Finally, suppose that we apply a field along $z$, the direction of the
spontaneous polarization. Here, for the first time in this discussion,
we have a non-zero linear contribution to the adiabatic temperature
change, and we can write $\Delta T(z)\approx \Delta T^{(1)}(z)+\Delta
T^{(2)}(z)$.

Concerning $\Delta T^{(2)}(z)$, the situation is identical to the
above cases for $T<T_{\rm C}$: it is dominated by $\pi^{(1)}_{zz} >
0$, which yields $\Delta T^{(2)}(z)<0$ regardless of the sign of the
applied field $E_{z}$.

In contrast, the linear contribution does depend on whether $E_{z}$ is
parallel or antiparallel to the spontaneous polarization. Further,
$\Delta T^{(1)}(z)$ is proportional to $\pi^{(0)}_{z}$, which is
negative when $P^{(0)}_{z}>0$. Hence, we have: $\Delta T^{(1)}(z)>0$
for a parallel field ($E_{z}>0$), and $\Delta T^{(1)}(z)<0$ when the
field goes against $P_{z}$.

Thus, we have two qualitatively different cases. If the applied field
goes against the polarization, $\Delta T^{(1)}(z)$ and $\Delta
T^{(2)}(z)$ are both negative, and we have every reason to expect
$\Delta T(z)<0$. However, for fields parallel to polarization, we have
a competition between the linear and quadratic contributions to
$\Delta T(z)$, and the net result is in principle
undetermined. Interestingly, experimental studies of FE phases show --
without exception, as far as we know -- that the temperature change is
positive for fields parallel to the spontaneous polarization
\cite{kutnjak15}, and negative for fields tending to reverse it
\cite{grunebohm18,thacher68,li13}. This is in agreement with the
expectations from our formalism, suggesting that, for the case of
parallel fields, the linear effect ($\Delta T^{(1)}(z) \propto
\pi^{(0)}_{z}E_{z} > 0$) dominates over the quadratic one ($\Delta
T^{(2)}(z) \propto \pi^{(1)}_{zz}E^{2}_{z} < 0$).

Hence, as summarized in Table~\ref{tab:summary}, we find that, in all
the cases considered, the leading non-zero contribution to $\Delta
T(\alpha)$ agrees in sign with the adiabatic temperature change
observed experimentally for relatively small applied
  fields.  We should note that the formalism just introduced bears
obvious similarities with previously proposed theories to discuss EC
effects, e.g. in the context of AFEs
\cite{pirc14,lisenkov16}. The novelty here relies on
  the fact that our equations are general and can be applied to any
  material and phase (FE, AFE, PE or simply dielectric), revealing the
  way they are connected and evidencing the (somewhat trivial) origin
  of the so-called inverse effects. Our formalism also emphasizes
that the basic EC response (i.e., the sign of the $T$-change) can be
understood from simple universal arguments, not relying on specific
atomistic or phenomenological models.

{\sl Numerical results}.-- To gain further insight, and to evaluate
the accuracy of low-order approximations to $\Delta T(\alpha)$, we now
compute explicitly the $\Delta T^{(n)}(\alpha)$ terms in
Eq.~(\ref{eq:newdeltaT}) for prototype compound PbTiO$_{3}$ (PTO).

We simulate PTO using the second-principles
(Refs.~\cite{wojdel13,garciafernandez16}; see
  Supp. Note~1) model potential first introduced in
Ref.~\onlinecite{wojdel13}, which has proven its accuracy in
reproducing the basic FE behavior of the material as well as many
subtle structural features, as e.g. related to its domain walls
\cite{wojdel14a,goncalves19}. Let us stress that, in
  these simulations, all the degrees of freedom for the lattice (i.e.,
  all atomic positions, all strains) are treated on equal footing;
  hence, our calculations include all contributions to the EC
  response, and there is in fact no easy way to differentiate them in
  the manner is often done in phenomenological Landau approaches
  (where it is natural to distinguish the ``dipole'' subsystem from
  the ``phonon'' bath \cite{kutnjak15}). Finally, let us mention that
the only noteworthy deficiency of our second-principles model pertains
to the predicted $T_{\rm C}$, which is lower than the experimental one
(510~K {\sl vs} 760~K), but this is not critical for the present
purposes.

We solve our PTO model as a function of temperature and applied
electric field \cite{wojdel13} by running Monte Carlo simulations,
using periodially-repeated supercells composed of $10\times 10\times
10$ or $12\times 12\times 12$ elemental perovskite units. (Larger
cells are considered in the proximity of $T_{\rm C}$.) At a given $T$,
we run 10,000 Monte Carlo sweeps for thermalization, followed by
75,000 to 100,000 sweeps to compute averages. We checked that these
calculation conditions yield sufficiently accurate results.

The key quantities we monitor are the equilibrium polarization ${\bf
  P}$, the dielectric susceptibility ${\boldsymbol \chi}$, the
pyroelectric vector ${\boldsymbol \pi}$, and the
specific heat $C_{\bf E}$, which we compute using standard
linear-response formulas \cite{garcia98}. (The formulas used here are
given in the Supp. Note~2; they closely resemble the ones employed in
other studies of EC effects by Monte Carlo simulations, as e.g. in
Refs.~\onlinecite{jiang17,jiang18}.) We can thus compute the Taylor
series for ${\boldsymbol \pi}$ (Eq.~(\ref{eq:pi})) and for $\Delta
T(\alpha)$ (Eq.~(\ref{eq:newdeltaT})). (In Supp. Note~3 and
Supp. Figs.~1, 2 and 3 we give extra details on the calculation of the
$\Delta T^{(n)}(\alpha)$ contributions.) Our results are summarized in
Fig.~\ref{fig:pto}.

We obtain a FE phase transition at $T_{\rm C} = 510$~K
(Fig.~\ref{fig:pto}(a)), marked by a near divergence of the
susceptibility (Fig.~\ref{fig:pto}(b)); this is characteristic of a
weakly first-order transformation, and is in qualitative agreement
with the experimental result \cite{lines-book1977}. Without loss of
generality, we choose ${\bf P}^{(0)}=(0,0,P^{(0)}_{z})$, with
$P_{z}^{(0)}>0$ below $T_{\rm C}$. Of note is the qualitative change
of the phase transition as we apply an electric bias:
for small applied field, the susceptibility peak
  reaches higher values, reflecting the fact that the transition
  becomes more continuous \cite{liu16}; for larger fields, the
transition becomes diffuse and the related anomalies in the
susceptibility (Fig.~\ref{fig:pto}(b)) and heat capacity
(Fig.~\ref{fig:pto}(c)) tend to disappear.

For clarity, Fig.~\ref{fig:pto}(b) only shows the temperature and
field dependence of one susceptibility component,
$\chi_{zz}$. However, note that, by symmetry, the ${\boldsymbol \chi}$
tensor only has diagonal components; further, for $T>T_{\rm C}$ they
are all equal (cubic phase), while for $T<T_{\rm C}$ we have
$\chi_{xx}=\chi_{yy}>\chi_{zz}$ (tetragonal phase). (In the FE phase,
the polar direction is electrically stiffer; this is a well-known
feature of FE perovskites, related to what is usually called {\sl easy
  polarization rotation} \cite{fu00}.) As can be seen in the
Supp. Fig.~4, the three components have an approximate Curie-Weiss
behavior, with a maximum at $T_{\rm C}$ and a monotonic $T$-dependence
on both sides of the transition point.

Figure~\ref{fig:pto}(d) shows our results for the low-order
pyroelectric coefficients. $\pi_{z}^{(0)}$ is null above the
transition point and negative below it, as expected. As for
${\boldsymbol \pi}^{(1)}$, we find the expected behavior as well: the
tensor has non-zero diagonal components at all temperatures, nearly
diverging at $T_{\rm C}$, and changing sign as the system goes through
the transition. Note that our numerical results for $\pi^{(0)}_{z}$
and $\pi^{(1)}_{zz}$ ratify the competition that occurs when a field
is applied parallel to $P^{(0)}_{z}$, as discussed above. Also, our
results for ${\boldsymbol \pi}^{(1)}$ reveal an essentially isotropic
tensor at all temperatures, even in the tetragonal phase. (The
tetragonal symmetry implies
$\pi^{(1)}_{xx}=\pi^{(1)}_{yy}\neq\pi^{(1)}_{zz}$; yet, in
Fig.~\ref{fig:pto}(d), a sizeable difference between the $xx$ and $zz$
components is found only for $T\lesssim T_{\rm C}$.)

Figure~\ref{fig:pto}(e) shows the computed $\Delta T(z)$ for various
positive fields, with $E_{z}$ up to 10~MV/m, as a function of
temperature. The results are obtained by evaluating the Taylor series
in Eq.~(\ref{eq:newdeltaT}) up to 5th order, which is enough to get a
converged $T$-change except in the immediate vicinity of the phase
transition. More specifically, Fig.~\ref{fig:pto}(f) shows that, even
for a large field of 10~MV/m, the EC effect at $T<T_{\rm C}$ is
dominated by the leading contribution $\Delta T^{(1)}(z)$,
higher-orders becoming significant only very close to $T_{\rm C}$ (see
the result at 490~K). In fact, as shown in Supp. Fig.~2, we find that
the leading contribution to the EC response -- i.e., $\Delta
T^{(1)}(z)$ below $T_{\rm C}$ and $\Delta T^{(2)}(z)$ above $T_{\rm
  C}$, respectively -- is usually a very good approximation of the
total effect.

Let us stress that in Figs.~\ref{fig:pto}(e) and \ref{fig:pto}(f) we
restrict ourselves to fields that are small enough so that no
first-order phase transition is induced. This is why we do not show
any data very close to $T_{\rm C}$, and why the fields considered for
$T>T_{\rm C}$ are relatively tiny. (Close to $T_{\rm C}$, the PE phase
is easily transformed into the polar one. See representative results
in Supp. Fig.~3.) This is consistent with our {\em perturbative}
approach to compute $\Delta T(\alpha)$, which is designed to describe
the properties of the continuously-deformed zero-field state. To treat
a first-order transition, one could split the integral for $\Delta
T(\alpha)$ in low- and high-field parts, using different perturbative
${\bf P}({\bf E})$ expansions for each of them. Additionally, one
should account for the latent heat associated to the discontinuous
transformation \cite{tishin99,giguere99,sun00}. The information to
tackle such situations is in principle available from our Monte Carlo
calculations. (We can compute latent heats from the thermal-averaged
internal energies \cite{zhong95a}.)  However, we should note that, for
treating first-order transitions, direct non-perturbative
computational approaches based on microcanonical molecular dynamics
\cite{marathe16} or constrained Monte Carlo simulations
\cite{ponomareva12} are better suited.

The values obtained for $\Delta T(z)$ (e.g., a maximum of 0.25~K when
$T_{\rm C}$ is approached from below for a field of 0.5~MV/m) are
comparable with EC effects measured for PTO; for example,
Ref.~\onlinecite{mikhaleva12} reports a temperature change of 0.1~K
for $T\lesssim T_{\rm C}$ and a field of 0.15~MV/m, and a maximum
$T$-change of 1.9~K at $T_{\rm C}$. (We also obtain $\Delta
S(z)\approx -2500$~J/Km$^{3}$ for a field of 0.5~MV/m at $T\lesssim
T_{\rm C}$, while Ref.~\onlinecite{mikhaleva12} reports a maximum
entropy change of about $-$16500~J/Km$^{3}$ for a field of 0.15~MV/m
applied exactly at $T_{\rm C}$.) Our results are also consistent with
other theoretical estimates of the EC effect for PTO
\cite{lisenkov13}. Finally, in Supp. Fig.~5 we evaluate the
approximations made in Eq.~(\ref{eq:newdeltaT}) -- i.e., the use of
the zero-field values for $T$ and $C_{\bf E}$, so they can be taken
out of the integral --, and find that their impact is negligible (even
for large fields) except very close to $T_{\rm C}$.

{\sl Discussion}.-- In view of the above, let us comment on the usual
intepretation of EC effects in terms of field-induced order or
disorder. For illustrative purposes, it is convenient to pay attention
to the isothermal entropy change, which, from Eqs.~(\ref{eq:deltaS})
and (\ref{eq:pi}), can be written as
\begin{equation}
  \begin{split}
    \Delta S(\alpha) = & \pi^{(0)}_{\alpha} E_{\alpha} +
    \frac{1}{2}\pi^{(1)}_{\alpha\alpha}E_{\alpha}^{2} + {\cal
      O}(E_{\alpha}^{3}) \\
    = & \Delta S^{(1)}(\alpha)+\Delta S^{(2)}(\alpha)+... \\
    = & \left(\pi^{(0)}_{\alpha} +
    \frac{1}{2}\pi^{(1)}_{\alpha\alpha}E_{\alpha} + ... \right)
    E_{\alpha} \; . \\
  \end{split}
 \label{eq:newdeltaS} 
\end{equation}
The last line suggests that we can think of $\Delta S(z)$ as depending
linearly on $E_{z}$, the corresponding proportionality constant having
spontaneous ($\sim \pi^{(0)}_{\alpha}$) and field-induced ($\sim
\pi^{(1)}_{\alpha\alpha}E_{\alpha}$) pyroelectric contributions.

Let us consider a FE state with $P^{(0)}_{z}>0$, and imagine we apply
an electric field $E_{z}>0$. For simplicity (and without loss of
generality), let us also assume that the dielectric response is
dominated by the linear effect $\chi^{(0)}_{zz}$, higher order terms
being negligible to a good approximation. In such a situation, it is
physically sound to assume that the applied field creates {\em order},
as it contributes to further align the local electric dipoles in the
FE state and results in a larger {\em order} parameter $P_{z}$. This
field-induced ordering is captured by $\chi^{(0)}_{zz}>0$, and we know
that the effect gets stronger as we approach $T\lesssim T_{\rm C}$. As
a result of this ordering, we expect the entropy to decrease in an
isothermal process ($\Delta S_{z} <0$), and the temperature to rise
($\Delta T_{z}>0$) if the process is adiabatic. These are clear
expectations that one would hardly question.

Let us inspect how these expected variations of $S$ and $T$ come about
in our formalism. Following the simplification mentioned above (linear
dielectric response), we can write $\Delta S(z) \approx \Delta
S^{(1)}(z) + \Delta S^{(2)}(z)$. (The discussion for $\Delta T(z)$ is
analogous.) The first term ($\Delta S^{(1)}(z) = \pi^{(0)}_{z}E_{z}$)
results in a reduction of the entropy, as we have $\pi^{(0)}_{z}<0$
for the spontaneous pyroelectric effect. This seems consistent with
the ordering argument given above. However, it is important to stress
that this term does {\em not} contain any information about the
dielectric response to the applied field, or about the order the field
may create. (There is no generally expected thermodynamic relationship
between $\partial P_{z}/\partial E_{z}$ and $\partial P_{z}/\partial
T$.) Instead, the entropy change is fully determined by the
$T$-dependence of the spontaneous polarization: $P_{z}^{(0)}>0$
decreases as we approach $T_{\rm C}$, a behavior that is {\sl normal}.

The response to an applied field does control the quadratic
contribution $\Delta S^{(2)}(z)$, where the spontaneous pyroelectric
effect in $\Delta S^{(0)}(z)$ is replaced by the field-induced one
($\sim \pi^{(1)}_{zz}E_{z}$). As mentioned above, it is clear that the
applied field strengthens the dipole order of the FE state, as
quantified by its order parameter: we have $P_{z}^{(0)}>0$ and a
field-induced polarization change
$\epsilon_{0}\chi^{(0)}_{zz}E_{z}>0$. However, the corresponding
entropy change is positive, since $\pi^{(1)}_{zz} \propto \partial
\chi^{(0)}_{zz}/\partial T >0$ for $T<T_{\rm C}$. Thus, the same
physical mechanism (captured by $\chi^{(0)}_{zz}$) results in
field-induced order {\em and} a positive contribution to the
entropy. Indeed, in what concerns the isothermal entropy change, what
matters is {\em not} the ordering of the dipoles at a given
$T$. Instead, we have to pay attention to how the field-induced effect
changes with temperature. In a FE state, the ordering caused by the
field grows as we heat up towards $T_{\rm C}$, that is, we have a
positive field-induced pyroelectric effect. This somewhat anomalous
behavior (the higher the temperature, the greater the -- induced --
order) is opposite to the spontaneous pyroelectric effect, and is the
one causing $\Delta S^{(2)}(z)>0$ (and $\Delta T^{(2)}(z)<0$).

Hence, the quadratic contribution to the EC effect in a FE state,
dominated by the linear susceptibility $\chi^{(0)}_{zz}$, seems
paradoxical: the field creates order (as quantified by the order
parameter) but the entropy increases. The key to the puzzle is that
polarization (order) created at a given $T$ is not what controls the
entropy; its $T$-derivative is. This point is hardly new, as it is
quite clear from the well-known equations governing EC effects
(Eqs.~(\ref{eq:deltaT}) and (\ref{eq:deltaS})); yet, it becomes
particularly apparent in our perturbative approach.

Keeping the above in mind, it is easy to interpret the EC effects in
PE and AFE states, which are dominated by the quadratic
contribution. In the PE case, the field-induced order decreases as $T$
increases: at $T>T_{\rm C}$ we have $\pi^{(2)}_{zz}<0$, which yields
$\Delta S(z)\approx\Delta S^{(2)}(z)<0$ and $\Delta T(z)\approx\Delta
T^{(2)}(z)>0$. It is tempting to interpret this result by focusing on
the response at a given $T$ (the field creates order, hence the
entropy should get reduced), but that would not be correct. Instead,
the focus should be on the field-induced pyroelectric effect which, in
this case, behaves in the {\em normal} way, i.e., we have a weaker --
induced -- order at higher temperature.

Finally, the {\em inverse} EC behaviour of AFE states is also readily
explained within this picture. In that case we are at $T<T_{\rm C}$,
and have $\Delta S(z)\approx\Delta S^{(2)}(z) \propto
(\pi^{(2)}_{zz}E_{z})E_{z}>0$ (and $\Delta T(z)\approx\Delta
T^{(2)}(z)<0$). According to the above discussion, this inverse
behavior stems from the fact that the applied field creates
polarization more efficiently as $T$ grows toward $T_{\rm
  C}$. Consequently, this disproves the frequent interpretation that
the inverse EC response of AFE states is caused by field-induced
disorder at a given $T$. As a matter of fact, the above analysis shows
that arguments focusing on the field-induced (dis)order at a given $T$
-- and overlooking the $T$-derivative -- are incorrect.

To conclude this part, let us comment on a FE type absent in our
discussion: relaxors. As mentioned above (Eq.~(\ref{eq:deltaT}),
ergodicity breaks down in these materials, and the use of the Maxwell
relation our formalism relies on is questinable. Nevertheless, some
experimental \cite{legoupil12} and theoretical \cite{jiang18} reports
suggest that, in fact, Eq.~(\ref{eq:deltaT}) approximately capures the
EC response of these systems at a quantitative level; further, it
seems to yield qualitatively correct results. Hence, we think that, in
the future, it may be worth considering whether our formalism, and the
kind of {\em rules} summarized in Table~\ref{tab:summary}, might be
useful in investigations of relaxors.

In summary, we have introduced a perturbative approach to the
electrocaloric effect. This formalism can be applied in quantitative
simulation studies, in a straightforward way as long as field-induced
(first-order) phase transitions are not present. More specifically,
our simulations for ferroelectric PbTiO$_{3}$ show that, except in the
vicinity of the phase transition, a low-order approximation captures
the electrocaloric response with great quantitative accuracy. Most
importantly, our formalism unifies and clarifies the physical
interpretation of all the small-field electrocaloric effects (normal
and inverse) observed in ferroelectric and antiferroelectric
materials.

We thank E.~Defay for many fruitful discussions. Work funded by the
Luxembourg National Research Fund through project
INTER/RCUK/18/12601980.

\end{document}